\documentclass[smallextended]{svjour2}                    
\smartqed  
\usepackage{graphicx}
\usepackage{graphicx}
\usepackage[dvipsnames,rgb,dvips]{xcolor}
\usepackage{psfrag}

\usepackage{amssymb,amsmath}
\newcommand{\ve}[1]{\ensuremath{\mbox{\boldmath$#1$}}}
\newcommand{\sve}[1]{\ensuremath{\mbox{\footnotesize\boldmath$#1$}}}
\newcommand{\ma}[1]{\ensuremath{\mathbb{#1}}}

\renewcommand{\Re}{\ensuremath{\operatorname{\mathfrak{Re}}}}

\newcommand\nn{\nonumber}

\newcommand{\ku}{\ensuremath{\mbox{Ku}}}

\newcommand{\st}{\ensuremath{\mbox{St}}}
\newcommand{\tr}{\ensuremath{\mbox{Tr}}}

\newcommand\transpose{^{\mathrm T}}

\newcommand{\eqnlab}[1]{\label{eq:#1}}

\newcommand{\eqnref}[1]{(\ref{eq:#1})}

\newcommand{\Eqnref}[1]{Eq.~(\ref{eq:#1})}

\newcommand{\Secref}[1]{Section~\ref{sec:#1}}

\journalname{Journal of Statistical Physics}
\begin{document}

\title{Lyapunov exponents for particles advected in compressible
random velocity fields at small and large Kubo numbers }

\author{K. Gustavsson        \and B. Mehlig\footnote{Corresponding author: {\tt Bernhard.Mehlig@physics.gu.se}, phone: +46 31 786 9170, fax: +46 31 772 2092}}
\institute{Department of Physics, Gothenburg University, 41296 Gothenburg, Sweden }
\date{Received: date / Accepted: date}
\maketitle

\begin{abstract}
We calculate the Lyapunov exponents describing spatial clustering of particles advected in
one- and two-dimensional random velocity fields at finite Kubo numbers $\ku$
(a dimensionless parameter characterising the correlation time of the velocity field).
In one dimension we obtain accurate results up to $\ku\sim 1$ by resummation of a perturbation expansion in $\ku$.
At large Kubo numbers we compute the Lyapunov exponent by taking into
account the fact that the particles follow the minima of the
potential function corresponding to the velocity field.
The Lyapunov exponent is always negative.
In two spatial dimensions the sign of the maximal Lyapunov exponent $\lambda_1$ may change,
depending upon the degree of compressibility of the flow and the Kubo number.
For small Kubo numbers we compute the first four non-vanishing terms in the small-$\ku$ expansion of the Lyapunov exponents.
By resumming these expansions we obtain a precise estimate of the location of the path-coalescence
transition (where $\lambda_1$ changes sign) for Kubo numbers up to approximately ${\rm Ku} = 0.5$.
For large Kubo numbers we estimate the Lyapunov exponents for a partially compressible velocity field by assuming that the particles
sample those stagnation points of the velocity field that have a negative real part of the maximal eigenvalue of the matrix of flow-velocity gradients.
\\
\\
{\em Keywords:} Advection, Compressible velocity fields, Clustering, Lyapunov exponents, Kubo number
\end{abstract}

\section{Introduction}
Consider many small tracer particles advected in a random or chaotic compressible flow.
An initially uniform scatter of particles cannot remain uniform because particles advected in smooth compressible flows
cluster together. An example of this effect is discussed by
Sommerer and Ott \cite{Som93} who describe experiments following fluorescent tracers floating
on the surface of an unsteady flow. Since the particles are constrained to the surface of the flow, they experience local up- and
down-welling regions as sources and sinks, rendering the surface flow compressible.
As a consequence the particles form fractal patterns. The authors of \cite{Som93} interpret these patterns in terms of random dynamical maps and estimate
the Lyapunov fractal dimension. This dimension is computed from the Lyapunov exponents by means of the Kaplan-Yorke formula \cite{Kap79}.
The Lyapunov exponents $\lambda_1>\ldots>\lambda_d$
(here $d$ is the spatial dimension) describe the long-term evolution of the patterns formed by the particles.
The maximal Lyapunov exponent $\lambda_1$
describes the dynamics of an initially infinitesimal separation between two particles.
When $\lambda_1<0$ separations between nearby particles must decrease
on the long run, clustering is strong.
This regime was referred to as \lq path-coalescence phase\rq{} in \cite{Wil03}.
The path-coalescence transition occurs at $\lambda_1=0$: when $\lambda_1>0$ separations typically grow, but clustering can nevertheless be substantial \cite{Wil12}.
The sum $\lambda_1+\lambda_2$ describes the evolution of
a small area element spanned by the separation vectors of three nearby particles, and so forth.

Refs.~\cite{Bof04,Cre04} summarise results of direct numerical simulations of tracers floating on the surface of turbulent flows,
and characterise the resulting fractal patterns in terms of their Lyapunov dimensions.

Another example is that of inertial particles in turbulent flows.
Finite inertia allows particles to detach from the flow. This effect gives rise to fractal patterns
of inertial particles suspended in incompressible flows \cite{Bec03b,Wil07}. When the particle inertia is small
it is commonly argued that the resulting fractal patterns can be understood in terms of a model that describes
particles advected in a slightly compressible particle-velocity field \cite{Bal01}.
The small correction term that renders the particle-velocity field compressible at
small particle inertia was first derived by Maxey \cite{Max87}.

This approach is frequently used in the literature to explain spatial clustering (so-called \lq preferential concentration\rq{}) of inertial particles suspended in
turbulent flows. We remark that this approach must fail when inertial effects become stronger. In this
case the particle-velocity field develops singularities (so-called \lq{}caustics\rq{}) that preclude
the existence of a smooth particle-velocity field (see e.g. \cite{Fal02,Wil05,Gus12,Gus13}). The singularities
give rise to large relative velocities between nearby particles \cite{Wil06,Gus11b,Gus13b}.

Many authors have studied the Lyapunov exponents of particles advected in turbulent, random, and chaotic velocity fields numerically.
Analytical results could only be derived in certain limiting cases though.
A limit that allows analytical progress in terms of diffusion approximations
is $\ku \rightarrow 0$. The \lq Kubo number\rq{} ${\rm Ku}=u_0\tau/\eta$
is a dimensionless measure of the correlation time $\tau$ of the fluctuations of the underlying velocity field,
$u_0$ is the typical speed of the flow and $\eta$ its correlation length.
In the limit of $\ku\rightarrow 0$ the problem of calculating
the Lyapunov exponents simplifies considerably. In this limit the flow causes many weakly correlated
small displacements of the particles and diffusion approximations can be used
to compute the exponents for random Gaussian flows \cite{LeJ85,Wil07}
and in the Kraichnan model \cite{Che98,Fal01,Bec04}. The exponents describe the fluctuations
of small separations between particles (much smaller than the correlation length or the Kolmogorov length $\eta$),
inertial-range fluctuations are not relevant in this limit, and thus the results for smooth random velocity fields and for the Kraichnan model
are equivalent when $\ku\rightarrow 0$. Lyapunov exponents for inertial particles
in the limit of $\ku\rightarrow 0$ were computed in Refs.~\cite{Wil03,Meh04,Dun05} in one, two,
and three spatial dimensions respectively.

Less is known about clustering at finite Kubo numbers where the particles have sufficient time to preferentially
sample the sinks of the underlying velocity field. This effect is important in the examples mentioned above, but
it is not captured by theories formulated in terms of diffusion approximations in the limit of $\ku \rightarrow 0$.
At large Kubo numbers the spatial patterns formed by the particles must depend
on the details of the fluctuations of the underlying flow, but it is not known how to analytically compute the Lyapunov exponents of particles advected in
compressible velocity fields at finite
Kubo numbers. We note however that an expression for the maximal Lyapunov exponent in incompressible two-dimensional
flows at finite Kubo numbers was obtained by Chertkov {\em et al.} \cite{Che96}.
Approximating the fluctuations of the flow-velocity gradient by telegraph noise with a finite
correlation time, Falkovich {\em et al.} computed Lyapunov exponents in one-dimensional
and incompressible two-dimensional models for advected and inertial particles \cite{Fal07a,Fal07}.
Also, Dhanagare {\em et al.} \cite{Mus13} recently investigated the spatial clustering of particles
advected in compressible random renovating flows. 

In this paper we compute the Lyapunov exponents
for particles advected in one- and two-dimensional compressible Gaussian random velocity fields with finite Kubo numbers
(the model is defined in Section \ref{sec:model}).  We use an approach recently developed to describe
incompressible turbulent aerosols at finite Kubo numbers \cite{Gus11}, generalising
a method used by Wilkinson \cite{Wil11} to compute the  Lyapunov exponent for particles advected in a one-dimensional random velocity field to lowest order in $\ku$.
This approach expresses the fluctuations of the flow-velocity gradient
along the particle trajectories at finite Kubo numbers in terms of correlation functions of the flow velocity
and its derivatives at fixed positions in space.  A perturbation expansion in $\ku$ is obtained by iteratively refining
an approximation for the paths taken by the particles \cite{Gus11}.

In Section \ref{sec:1d} we develop perturbation series to order $\ku^{12}$ in one spatial dimension.
By comparison with computer simulations we show that a Pad\'e{}-Borel resummation of the series yields accurate results
up to $\ku \sim 1$.  For $\ku\gg 1$ the resummation fails. In this case
the particles are predominantly found near stagnation points of the flow (where the flow velocity
vanishes) with negative velocity gradients, that is near the minima of the corresponding potential function.
In this regime the Lyapunov exponent is determined by the flow-gradient fluctuations near these points,
and we show how to compute the exponent using the Kac-Rice formula \cite{Kac43,Ric45} for counting
singular points of random functions.

Section~\ref{sec:2d} summarises the corresponding results for two-dimensional velocity fields.
For small Kubo numbers we compute the first four non-vanishing terms in a perturbation
expansion (to order $\ku^8$). We compare the results
of a Pad\'e{}-Borel resummation of this series with results of numerical simulations.
We find that resummation of the perturbation series provides an accurate estimate
of the location of the path-coalescence transition (the degree of compressibility
where the maximal Lyapunov exponent $\lambda_1$ changes sign) for Kubo numbers up
to $\sim 0.5$.

For much larger Kubo numbers, particles in a purely compressible velocity field (that can be written as
the gradient of a potential function)  spend most of their time near the minima of the potential function,
that is near stagnation points of the velocity field with 
negative real part of the maximal eigenvalue of the matrix of flow-velocity gradients.
As in the one-dimensional case the Lyapunov exponents can be computed using the Kac-Rice formula.
Our results agree well with those of numerical simulations of particles in velocity fields with a compressible component, at large but finite Kubo numbers.

Section \ref{sec:conc} summarises our conclusions.

\section{Model}
\label{sec:model}
\begin{table}[t]
\begin{tabular}{ccccc}
\hline\hline\\
& $\beta^2$ & $\Gamma$ & $\Delta_C$ & $\wp$ \\[1mm]
\hline\\[-2mm]
$\beta^2$ & -                                    & $\frac{d+1+\Gamma(1-d)}{3\Gamma-1}$   & $\frac{(d-1)\Delta_C}{4+2d-3\Delta_C}$ & $\frac{\wp(1-d)}{\wp-1}$ \cr
$\Gamma$ & $\frac{d+1+\beta^2}{d-1+3\beta^2}$   & -  & $\frac{d+1-\Delta_C}{d-1}$   & $\frac{d+1-2\wp}{(d-1)(1+2\wp)}$    \cr
$\Delta_C$ & $\frac{2(d+2)\beta^2}{d-1+3\beta^2}$ & $d+1+(1-d)\Gamma$                     & -  & $\frac{2(d+2)\wp}{1+2\wp}$ \cr
$\wp$ & $\frac{\beta^2}{d-1+\beta^2}$ & $\frac{d+1+\Gamma(1-d)}{2(1+\Gamma(d-1))}$ & $\frac{\Delta_C}{2(d+2-\Delta_C)}$ &  -  \cr\\
\hline\hline
\end{tabular}
\caption{\label{tab:1} Conversion table comparing different parametrisations of compressibility in $d$-dimensional random velocity fields.
The parameters $\beta$ and $\Gamma$ were introduced in Ref.~\cite{Meh04} in two and in Ref.~\cite{Meh05} in three
spatial dimensions. The parameter $\Delta_C$ in Eq.~(6.21) in Ref.~\cite{Gus09lic}, and $\wp$ in Eq.~(57) in Ref.~\cite{Fal01} (see also Ref.~\cite{Bec04}).}
\end{table}

We study particles advected in one- and two-dimensional
Gaussian random velocity fields. In one dimension the
equation of motion is
\begin{equation}
\label{eq:1d}
\dot x_t = u(x_t,t)\,.
\end{equation}
Here $x_t$ denotes the position of the particle at time $t$, the dot denotes
a time derivative, and $u(x,t)$ is a Gaussian random velocity field. We write
$u=u_0\partial\psi/\partial x$ where $u_0$ is the typical speed of the flow, and $\psi(x,t)$ is a Gaussian random function
with zero mean values and correlation function
\begin{equation}
\label{eq:1dcorr}
\langle \psi(0,0)\psi(x,t)\rangle = e^{-x^2/(2\eta^2)-|t|/\tau}\,.
\end{equation}
The correlation length is denoted by $\eta$, and the correlation time is denoted by $\tau$.
In two spatial dimensions we write
\begin{align}
\dot{\ve x}_t&=\ve u(\ve x_t,t)
\eqnlab{eqm_r}
\end{align}
with $\ve x = (x,y)\transpose$. The velocity field is defined as \cite{Meh04}:
\begin{equation}
\label{eq:def1}
\ve u=\frac{u_0}{\sqrt{2(1+\beta^2)}}[\ve\nabla\phi\wedge\hat{\ve e}_z+\beta\ve\nabla\psi]\,,
\end{equation}
where $\hat{\ve e}_z$ is the unit vector in the $z$-direction and where $\psi$ and $\phi$ are independent Gaussian random functions
with zero means and correlation functions
\begin{align}
\label{eq:def2}
\langle\psi(\ve x,t)\psi(\ve 0,0)\rangle=\langle\phi(\ve x,t)\phi(\ve 0,0)\rangle={\rm e}^{-|\sve x|^2/(2\eta^2)-|t|/\tau}\,.
\end{align}
The first term in Eq.~(\ref{eq:def1}) is an incompressible (or \lq solenoidal\rq{}) contribution. The second term
is a compressible (or \lq potential\rq{}) contribution.

As in one dimension the speed-, length-, and time
scales of the flow are denoted by $u_0$, $\eta$ and $\tau$. The Kubo number is given by $\ku = u_0 \tau/\eta$.
In the following we adopt dimensionless units $t=\tau t'$, $x=\eta x'$, $u=u_0 u'$ and we drop the primes.
A second dimensionless parameter of the problem is the degree of compressibility. Ref.~\cite{Meh04}
introduced the parameter
\begin{equation}
\Gamma = \frac{3+\beta^2}{3\beta^2+1}\,.
\end{equation}
It ranges from $1/3$
($\beta \rightarrow \infty$, compressible)
to $\Gamma=3$
($\beta = 0$, incompressible).
In the limit of $\ku\rightarrow 0$, the maximal Lyapunov exponent is negative for $\Gamma \leq 1$ ($\beta \geq 1$) and positive otherwise.
Other authors parametrise the degree of compressibility in other ways. Table~\ref{tab:1} compares different definitions.

\section{One spatial dimension}
\label{sec:1d}
The Lyapunov exponent
\begin{eqnarray}
\label{eq:l1d}
\lambda &=& \lim_{t\rightarrow \infty} \frac{1}{t} \log \Big| \frac{\delta x_t}{\delta x_0}\Big|
\end{eqnarray}
describes the long-term growth (or decline) of the separation $\delta x_t$ between two
initially infinitesimally close particles. It
is computed by linearising the equation of motion (\ref{eq:1d})
to find the dynamics of a small separation $\delta x_t$ between two neighbouring particles. Using the dimensionless
variables introduced in Section \ref{sec:model} we have:
\begin{equation}
\frac{{\rm d}\delta x_t}{{\rm d}t} = {\rm Ku}\, \frac{\partial u}{\partial x}(x_t,t)\, \delta x_t\equiv \ku\,A(x_t,t)\,\delta x_t\,.
\eqnlab{eqm_x1d}
\end{equation}
Here the flow-velocity gradient at position $x$ at time $t$ is denoted by $A(x,t)$. It follows that the Lyapunov exponent is given by
by the average flow-velocity gradient evaluated at the particle position $x_t$:
\begin{equation}
\lambda = \ku\,\lim_{t\rightarrow \infty} \langle A(x_t,t)\rangle\,,
\eqnlab{lambda1_1d}
\end{equation}
where $\langle \cdots \rangle$ denotes an average over flow realisations.

\begin{figure}[t]
\includegraphics[width=5cm]{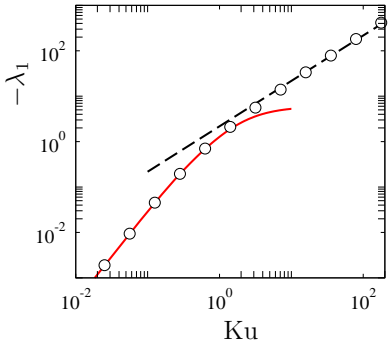}
\hspace*{2mm}
\caption{\label{fig:1} ({\em Online color}).
Lyapunov exponent in one spatial dimension from direct numerical simulations of the
model described in \Secref{model} (symbols), theory for large $\ku$ (\ref{eq:lambda_largeKu_1D}), dashed line, and
Pad\'e{}-Borel resummation (\ref{eq:pdbt}) of the perturbation series for small $\ku$ (solid red line).  }
\end{figure}

The Lyapunov exponent is computed
by expanding the implicit solution of (\ref{eq:1d}).
In  dimensionless units it is given by:
\begin{equation}
x_t-x_0=\ku\int_{0}^{t} {\rm d}t'u(x_{t'},t')\equiv \xi_t\,.
\label{eq:rsolution}
\end{equation}
Here $x_0$ is the initial particle position, and $\xi_t=x_t-x_0$ is the difference between the trajectory of a particle
and its initial position (to be distinguished from the separation $\delta x_t$ between two neighbouring particles at time $t$).
Since $\xi_t$ is proportional to $\ku$ it can be considered small provided that $\ku$ is sufficiently small.  In this case we expand  $u(x_t,t)$ in powers of $\xi_t$:
\begin{equation}
u(x_t,t)=
\sum_{n=0}^\infty \frac{1}{n!} \frac{\partial^n u}{\partial x^n}(x_0,t) \,\xi_t^n\,.
\label{eq:uexpansion}
\end{equation}
Inserting $\xi_t = x_t-x_0$ from Eq.~(\ref{eq:rsolution}) into  Eq.~(\ref{eq:uexpansion}) 
and iterating Eq.~(\ref{eq:uexpansion}) yields a perturbation series
of $u(x_t,t)$ in terms of powers of $\ku$. In the same way an expansion of $A(x_t,t)$ is found. To third order in
$\ku$ we obtain for example:
\begin{align}
A(x_t,t)&=A(t)
+\ku\, \partial_x^2u(t)\int_0^t{\rm d}t_1 u(t_1)\nn\\
&+\ku^2\Big[
\frac{1}{2} \partial_x^3u(t) \int_0^t{\rm d}t_1\int_0^t{\rm d}t_2 u(t_1) u(t_2) \nn\\
&\hspace*{1cm}+ \partial_x^2u(t) \int_0^t{\rm d}t_1 \int_0^{t_1}{\rm d}t_2 A(t_1)u(t_2)
\Big]
\nn\\&
+\ku^3\bigg[
\frac{1}{2}\partial_x^2u(t) \int_0^t{\rm d}t_1 \int_0^{t_1}{\rm d}t_2 \int_0^{t_1}{\rm d}t_3 u(t_2) u(t_3) \partial_x^2u(t_1) \nn\\
&\hspace*{1cm}+
\frac{1}{2}\partial_x^3u(t) \int_0^t{\rm d}t_1 \int_0^t{\rm d}t_2 \int_0^{t_1}{\rm d}t_3 A(t_1)u(t_2) u(t_3) \nn
\\&\hspace*{1cm}+
\frac{1}{2}\partial_x^3u(t) \int_0^t{\rm d}t_1 \int_0^t{\rm d}t_2 \int_0^{t_2}{\rm d}t_3 A(t_2)u(t_1) u(t_3)\nn\\
&\hspace*{1cm}  +\partial_x^2u(t)\int_0^t{\rm d}t_1 \int_0^{t_1}{\rm d}t_2 \int_0^{t_2}{\rm d}t_3 u(t_3)A(t_1) A(t_2)\nn\\
&\hspace*{1cm}  + \frac{1}{6}\partial_x^4u(t)\int_0^t{\rm d}t_1 \int_0^t{\rm d}t_2 \int_0^t{\rm d}t_3 u(t_1) u(t_2) u(t_3)
\bigg]\,,
\end{align}
where $u(t) \equiv u(x_0,t)$, $A(t) \equiv A(x_0,t)$, and so forth.
Averaging yields an expression for the Lyapunov exponent in terms of time integrals of Eulerian correlation functions of the velocity field
and its derivatives. Evaluating these correlation functions
requires computing averages of products of $u(t)$ and its spatial derivatives 
$\partial_x^k u(t)\equiv\partial_x^k u(x,t)|_{x=x_0}$ (for $k=1,2,\ldots$) evaluated at different times.
For a Gaussian random velocity field we use Wick's theorem which states that
the average of a product of $n$ Gaussian variables $z_1,\dots z_n$ is equal to the
sum of all ways of decomposing the product into a products of covariances.
The averaged product $\langle z_1,\dots z_n\rangle$  is  calculated using the known covariances
$\langle z_iz_j\rangle$. Averages of products of an odd number of factors vanish when $\langle z_i\rangle = 0$.
In one spatial dimension, the covariances of the velocity and its spatial derivatives are determined by Eq.~(\ref{eq:1dcorr}):
\begin{align}
&\left\langle\frac{\partial^m u}{\partial x^m}(x_0,t_1)\frac{\partial^n u}{\partial x^n}(x_0,t_2)\right\rangle\nn\\
&\hspace*{1cm}=
\left\{
\begin{array}{ll}
(-1)^{(n-m)/2}(m+n+1)!!e^{-|t_1-t_2|} & \mbox{if }m+n\mbox{ even}\cr
0 & \mbox{otherwise}
\end{array}
\right.
\label{eq:correlation_uu_1d}
\end{align}
with $m,n=0,1,2,\dots$. In this way we obtain an expansion of the Lyapunov
exponent in powers of $\ku$.  The final result to order $\ku^{12}$ is:
\begin{align}
&\lambda=-3\ku^2+12\ku^4-\frac{269}{2}\ku^6+2324\ku^8\nn\\
&\hspace*{1cm}-\frac{1759529}{20}\ku^{10}+\frac{1072147807}{720}\ku^{12}+\cdots\,.
\eqnlab{lambda_smallKu_1D}
\end{align}
This series expansion is asymptotically divergent: it diverges for any fixed value of $\ku$ but every partial sum of the series approaches $\lambda$
as $\ku\rightarrow 0$. At large orders $k$ the coefficients $c_k$ in the series (\ref{eq:lambda_smallKu_1D})  are of the form \cite{boyd}
\begin{align}
c_k\sim a S^{-k}(k-1)!(1+b/k+\cdots )
\label{eq:alfak_ansatz}
\end{align}
with $S$ close to $1/6$, $a\approx 0.23$, and $b \approx 1$.
The series (\ref{eq:lambda_smallKu_1D}) can be resummed
by Pad\'e-Borel resummation \cite{boyd}. The result is expressed
as the Laplace transform of the
so-called \lq Borel sum' (assumed to have
a finite radius of convergence due to the extra factor of $1/k!$):
\begin{equation}
B(\ku^2) = \sum_{k=1}^\infty \frac{c_k}{k!} \ku^{2k}\,.
\end{equation}
The Lyapunov exponent is estimated by
\begin{equation}
\label{eq:pdbt}
\lambda = \Re\int_C {\rm d}s \,{\rm e}^{-s} B(\ku^2 s)\,.
\end{equation}
The integration path $C$ is taken to be a ray in the upper right
quadrant of the complex plane. In order to compute the integral the Borel
sum must be analytically continued outside its radius of convergence.
This can be achieved by \lq Pad\'e approximants' \cite{Ben78}.
We know $6$ non-zero coefficients in the sum, this allows us
to compute the Pad\'e{} approximant of third orders in $\ku^2$ in numerator and denominator:
\begin{eqnarray}
\label{eq:b33}
B_{[3,3]}({\rm Ku}^2) &=-
\frac{\frac{14440189013 \,{\rm Ku}^6}{957721800}+\frac{985684759 \,{\rm Ku}^4}{63848120}+3 \,{\rm Ku}^2}{\frac{88699515137 \,{\rm Ku}^6}{34477984800}+\frac{11344821011\, {\rm Ku}^4}{957721800}+\frac{1368773479 \,{\rm Ku}^2}{191544360}+1}\,.
\end{eqnarray}
Eqs.~(\ref{eq:pdbt}) and (\ref{eq:b33}) together determine an approximation for the Lyapunov exponent.
The corresponding result is shown in Fig.~\ref{fig:1}, compared with results of
numerical simulations of the model. We observe good agreement for Kubo numbers up to order unity.
If more coefficients in the perturbation series were known, higher-order Pad\'e{} approximants could be computed
to improve the accuracy of the Pad\'e-Borel resummation.

For large values of $\ku$ the resummation fails.
We now show how to approximate the Lyapunov exponent at large but finite values of $\ku$
(the limit $t \rightarrow \infty$ in Eq.~(\ref{eq:l1d}) is taken at a finite value of $\ku$).
At large Kubo numbers the particles spend most of their time in the vicinity of
the minima of the \lq potential\rq{} $V(x,t) = -\psi(x,t)$.
An approximate expression for the Lyapunov exponent can be obtained by
averaging the gradient $A$ at the minima, that is at the stagnation
points of the flow velocity $u$ with $A<0$.
The distribution of $A$ at $u=0$ can be estimated using
the Kac-Rice formula \cite{Kac43,Ric45}:
\begin{equation}
p(A) = \int {\rm d} A' |A'| \delta (A-A') P[0,A']\,.
\end{equation}
Here $P[u,A]$ is the joint distribution of the velocity field $u$ and its gradient $A$.
It is determined by the correlation function of $\psi$ given in Section~\ref{sec:model}. We find:
\begin{align}
p(A) =\frac{1}{6}|A|e^{-A^2/6}\,.
\end{align}
Since $p$ is symmetric in $A$, the distribution of negative values of $A$ at $u=0$ is given by $2\,p(A)$.
The Lyapunov exponent is thus given by
\begin{align}
\lambda=2\,\ku\int_{-\infty}^0\!\!\!{\rm d}A\, A\, p(A)=-\sqrt{\frac{3\pi}{2}}\ku\,.
\eqnlab{lambda_largeKu_1D}
\end{align}
The limiting behaviour (\ref{eq:lambda_largeKu_1D})  is shown in Fig.~\ref{fig:1}. It is in good
agreement with the results of numerical simulations.

\section{Two spatial dimensions}
\label{sec:2d}
\subsection{Small-$\ku$ limit}
Consider first the case of small Kubo numbers.
Now there are two Lyapunov exponents to compute, describing
the time evolution of the distance $|\delta\ve x_t|$ between two neighbouring
particles and of the infinitesimal area element $\delta {\cal A}_t$ spanned by
the separation vectors between three neighbouring particles:
\begin{eqnarray}
\lambda_1 &=& \lim_{t\rightarrow \infty} \frac{1}{t} \log \Big| \frac{\delta\ve x_t}{\delta\ve x_0}\Big|\,,\\
\lambda_1+\lambda_2 &=& \lim_{t\rightarrow \infty} \frac{1}{t}\log  \Big| \frac{\delta {\cal A}_t}{\delta {\cal A}_0}\Big|\,.
\eqnlab{lambda_2d_def}
\end{eqnarray}
As in the one-dimensional case, these Lyapunov exponents are computed by
linearising the equation of motion, Eq.~\eqnref{eqm_r} in this case.
The dynamics at small separations $\delta\ve x$ between two neighbouring particles is
\begin{equation}
\frac{{\rm d}}{{\rm d}t} {\delta\ve x}_t = \ku\,\ma A(\ve x_t,t)\,\delta\ve x_t\,.
\eqnlab{eqm_RR}
\end{equation}
Here $\ma A$ is the flow-gradient matrix with elements $A_{ij}\equiv\partial u_i/\partial x_j$.
The Lyapunov exponents \eqnref{lambda_2d_def} are calculated from
\begin{align}
\label{eq:l1}
\lambda_1 &= \ku\,\lim_{t\rightarrow \infty} \langle {\ve n}_t\transpose\,\ma A(\ve x_t,t)\,{\ve n}_t\rangle\,,\\
\lambda_1+\lambda_2 &= \ku\,\lim_{t\rightarrow \infty} \langle\tr\,\ma A(\ve x_t,t)\rangle\,.
\label{eq:l2}
\end{align}
These relations are analogous to the one-dimensional Eq.~(\ref{eq:lambda1_1d}).  In Eq.~(\ref{eq:l1}),
the vector ${\ve n}_t\equiv\delta\ve x_t/|\delta\ve x_t|$ is a time-dependent unit vector aligned with the separation vector between the two particles.
Its dynamics follows from \Eqnref{eqm_RR}:
\begin{align}
\dot{{\ve n}}_t &= \ku\,\left[\ma A(\ve x_t,t)\,{\ve n}_t - ({\ve n}_t\transpose\,\ma A(\ve x_t,t)\,{\ve n}_t)\,{\ve n}_t\right]\,.
\eqnlab{eqm_nn}
\end{align}
The expressions \eqnref{eqm_RR} -- \eqnref{eqm_nn} can be expanded analogously to the one-di\-men\-sio\-nal case
described in the previous section.  For the model flow given by Eqs.~(\ref{eq:def1}) and (\ref{eq:def2}) we
find to order $\ku^8$:
\begin{align}
\lambda_1&=\ku^2\frac{1-\beta^2}{1+\beta^2}-\ku^4\frac{6+7\beta^2-\beta^4}{(1+\beta^2)^2}+\ku^6\frac{423+972\beta^2+597\beta^4+20\beta^6}{6(1+\beta^2)^3}\nn\\
         &-\ku^8\frac{164136+521517\beta^2+591081\beta^4+255803\beta^6+19927\beta^8}{144(1+\beta^2)^4}
\eqnlab{lambda_smallKu_2D}
\end{align}
and
\begin{align}
\lambda_1+\lambda_2&=
-\ku^2\frac{4\beta^2}{1+\beta^2}
+2\ku^4\beta^2\frac{3+5\beta^2}{(1+\beta^2)^2}
-2\ku^6\beta^2\frac{33+123\beta^2+104\beta^4}{3(1+\beta^2)^3}
\nn\\&
+\ku^8\beta^2\frac{8718+51303\beta^2+92196\beta^4+51787\beta^6}{72(1+\beta^2)^4}\,.
\label{eq:sum_lambda_smallKu_2D}
\end{align}
The lowest order, $\ku^2$, is consistent with
the results quoted in \cite{LeJ85,Che98,Fal01,Bec04,Wil07}:
the maximal Lyapunov exponent changes sign at $\beta_{\rm c}=1$.
For $\beta=0$, Eqs.~(\ref{eq:lambda_smallKu_2D}) and (\ref{eq:sum_lambda_smallKu_2D})
yield the Lyapunov exponents for particles advected
in a two-dimensional incompressible Gaussian random  velocity field with finite
Kubo number: $\lambda_1+\lambda_2=0$ (there cannot be clustering of particles advected
in an incompressible flow), and Eq.~(\ref{eq:lambda_smallKu_2D})  corresponds to the advective
limit of Eq.~(8) in Ref.~\cite{Gus11}
(the limit of zero Stokes number, $\st\rightarrow 0$,
must be taken in this equation describing the
the maximal Lyapunov exponent of inertial particles).
The series   (\ref{eq:lambda_smallKu_2D}) and (\ref{eq:sum_lambda_smallKu_2D})
are expected to be asymptotically divergent.  Just as the one-dimensional result (\ref{eq:lambda_smallKu_1D}),
Eqs.~(\ref{eq:lambda_smallKu_2D}) and (\ref{eq:sum_lambda_smallKu_2D})
are expected to fail at large Kubo numbers.

The series (\ref{eq:lambda_smallKu_2D}) and (\ref{eq:sum_lambda_smallKu_2D}) can be resummed
as in the one-dimensional case.
\begin{figure}
\includegraphics[width=10cm]{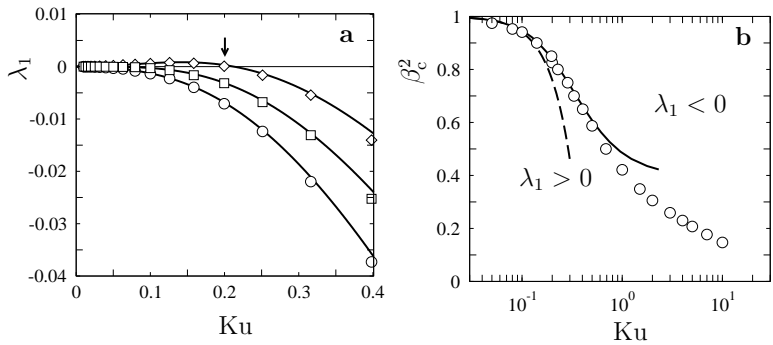}
\caption{\label{fig:3} {\bf a} Numerical results for maximal Lyapunov exponent for three different
compressibilities (see Table \ref{tab:1}): $\beta = 0.91$ ($\Diamond$), $\beta = 1$ ($\Box$),
and $\beta=1.11$ ($\circ$). For $\beta = 0.91$ the location of the path-coalescence transition
is indicated by an arrow. Also shown are results of Pad\'e{}-Borel resummations of the
series (\ref{eq:lambda_smallKu_2D}) [solid lines].
{\bf b} Shows location of the path-coalescence transition in the $\ku$-$\beta^2$ plane.
Results of numerical simulations ($\circ$), resummation of the perturbation
series (\ref{eq:lambda_smallKu_2D}) [solid line]. The dashed line corresponds to Eq.~(\ref{eq:bc}). }
\end{figure}
The results are seen  in Fig.~\ref{fig:3}. Panel {\bf a} shows results
for the maximal Lyapunov exponent for $\beta$ close to $\beta_{\rm c}=1$
(the location of the path-coalescence transition in the limit of $\ku \rightarrow 0$).
We see that at finite Kubo numbers the path-coalescence transition occurs
at $\beta_{\rm c}(\ku) < 1$. Comparing the first two terms in Eq.~(\ref{eq:lambda_smallKu_2D})
shows that to order $\ku^2$:
\begin{equation}
\label{eq:bc}
\beta_{\rm c}^2 = 1-6 \ku^2\,.
\end{equation}
For very small values of $\ku$ this agrees with the numerical results
in Fig.~\ref{fig:3}{\bf b}. This panel shows the location of the path-coalescence transition
in the $\ku$-$\beta^2$ plane.  At large values of $\ku$, Pad\'e{}-Borel resummations of the perturbation series (\ref{eq:lambda_smallKu_2D})
substantially improve the result.  We observe good agreement between the numerical results and those
of the resummation for values of $\ku$ up to approximately $0.5$.

The results summarised in Fig.~\ref{fig:3}  show that at larger Kubo numbers less compressibility is needed
to turn the maximal Lyapunov exponent negative. This is consistent with the behaviour observed at very
large Kubo numbers: in the following section we show that the particles preferentially sample
the attracting stagnation points of the velocity field in this limit. The contribution of these points increases
as the Kubo number becomes larger.
The observation that the effect of the compressible part of the velocity field is amplified at
large Kubo numbers is consistent with numerical results in random renovating flows \cite{Mus13}.
The dynamics of particles advected on the surface of a turbulent flow, by contrast, show a different behaviour \cite{Bof04}.
In this case it is observed that the path-coalescence transition occurs at larger values of the compressibility
for larger Kubo numbers. It would be of interest to determine which particular property of the turbulent
flow gives rise to this effect.

\subsection{Large-${\rm Ku}$ limit}
The large-$\ku$ limit in two-dimensional compressible flows is solved as in one spatial dimension.
The required distribution $p(\ma A)$ of the flow-gradient matrix $\ma A$ at $\ve u=0$
is found to be:
\begin{align}
p(\ma A)=\frac{\sqrt{(3+\beta^2)(1+3\beta^2)}}{\pi^2\beta}|\det\ma A|e^{-\sve a\transpose\ma C^{-1}\sve a/2}\,,
\end{align}
where $\ve a=(A_{11},A_{12},A_{21},A_{22})$ and
\begin{align}
\ma C=
\frac{1}{2(1+\beta^2)}\begin{pmatrix}
1+3\beta^2 & 0 & 0 & \beta^2-1\cr
0 & 3+\beta^2 & \beta^2-1 & 0\cr
0 & \beta^2-1 & 3+\beta^2 & 0\cr
\beta^2-1 & 0 & 0 & 1+3\beta^2
\end{pmatrix}\,.
\end{align}
The above expressions correspond to flows with both potential and solenoidal components
because we expect that the expressions for the Lyapunov exponents derived below are
not only valid for purely potential flows, but also yield good estimates for flows with a small solenoidal
component.

We change coordinates $s_\pm=(A_{11}\pm A_{22})/2$ and $u_\pm=(A_{12}\pm A_{21})/2$ to obtain
\begin{align}
p(s_+,s_-,u_+,u_-)&=\frac{\sqrt{(3+\beta^2)(1+3\beta^2)}}{4\pi^2\beta}|s_+^2+u_-^2-u_+^2-s_-^2|\nn\\
&\times \exp\left[-\frac{1+\beta^2}{2\beta^2}s_+^2-s_-^2-u_+^2-\frac{1+\beta^2}{2}u_-^2\right]\,.
\end{align}
The Lyapunov exponents are given by the eigenvalues of the strain matrix $\sigma_\pm=s_+\pm\sqrt{s_-^2+u_+^2-u_-^2}$ at the zeroes
of $\ve u$ subject to the constraint $\Re\,\sigma_\pm<0$.  This condition is equivalent to the condition $\tr\ma A<0$ and $\det\ma A>0$
and can be expressed as
\begin{equation}
s_+<-\Re \sqrt{s_-^2+u_+^2-u_-^2}\,.
\end{equation}
We have:
\begin{align}
\lambda_{1,2}&=4\ku
\int_{-\infty}^\infty{\rm d}s_+{\rm d}s_-{\rm d}u_+{\rm d}u_-\, \Theta\Big(-\Re\sqrt{s_-^2+u_+^2-u_-^2}-s_+\Big)\nn\\
&\times \Big(s_+\pm\sqrt{s_-^2+u_+^2-u_-^2}\Big)p(s_+,s_-,u_+,u_-)\,.
\end{align}
The factor $4$ is a normalisation factor due to the fact that we only consider matrices $\ma A$ with $\Re\,\sigma_\pm<0$. Further $\Theta(z)$ takes the value $\Theta(z)=1$ when $z>0$, and zero otherwise. To evaluate the integral we change variables according to
\begin{align}
&\left\{
\begin{array}{l}
s_-=w\sqrt{1+\xi^2}\cos\varphi\cr
u_+=w\sqrt{1+\xi^2}\sin\varphi\cr
u_-=w\xi
\end{array}
\right.\,
\mbox{ for } \Delta=s_-^2 + u_+^2 - u_-^2\ge 0
\\
&\left\{
\begin{array}{l}
s_-=w\xi\cos\varphi\cr
u_+=w\xi\sin\varphi\cr
u_-=w\sqrt{1+\xi^2}
\end{array}
\right.\,
\mbox{\hspace*{.83cm} for } \Delta=s_-^2 + u_+^2 - u_-^2<0
\end{align}
with $0\le w<\infty$, $-\infty<\xi<\infty$ and $0\le\phi<2\pi$.
Upon integrating over $\varphi$ and $\xi$ we obtain:
\begin{align}
\lambda_{1,2}&=
2\ku\frac{\sqrt{1+3\beta^2}}{\sqrt{2\pi}\beta}\int_{-\infty}^0{\rm d}s_+\int_{0}^\infty\!\!\!{\rm d}w\, w\,
\exp\Big[-\frac{\scriptstyle 1+\beta^2}{\scriptstyle 2\beta^2}s_+^2\Big]\\
&\times \Big\{
\Theta\Big(-\omega-s_+\Big)(s_+\mp\sqrt{w^2})|s_+^2-w^2|e^{-w^2}\nn\\
& +(s_+\mp\sqrt{-w^2})|s_+^2+w^2|{\rm erfc}\Big[\frac{\sqrt{\scriptstyle 3+\beta^2}w}{\scriptstyle \sqrt{2}}\Big]e^{w^2}
\Big\}\,.\nn
\end{align}
Performing the remaining integrals we find an approximation
for the Lyapunov exponents:
\begin{align}
\lambda_{1,2}&=\frac{\ku}{2(1+\beta^2)^{5/2}\sqrt{1+3\beta^2}\sqrt{\pi}}
\Big[
\sqrt{2}\beta(1+\beta^2)^{3/2}(-2\mp 3+\beta^2(-2\mp 5))\nn\\
&\hspace{2.3cm}-2\sqrt{2}\beta^3\sqrt{3+\beta^2}(1+3\beta^2)
\nn\\&
\hspace{2.3cm}\mp(1+\beta^2)(3+\beta^2)(1+3\beta^2)\arctan\Big(\sqrt{\frac{\scriptstyle 2\beta^2}{\scriptstyle 1+\beta^2}}\Big)
\Big]\,.
\eqnlab{lambda_largeKu_2D}
\end{align}
The sum of the Lyapunov exponents is given by
\begin{align}
\lambda_1+\lambda_2&=
-\frac{2\sqrt{2}\beta\ku}{(1+\beta^2)^{5/2}\sqrt{1+3\beta^2}\sqrt{\pi}}
\bigg[
(1+\beta^2)^{5/2}
+\beta^2\sqrt{3+\beta^2}(1+3\beta^2)
\bigg]\,.
\label{eq:sum_lambda_largeKu_2D}
\end{align}
\begin{figure}[t]
\includegraphics[width=10cm]{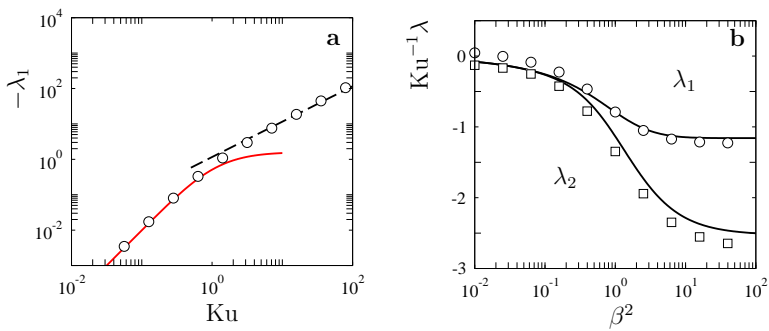}
\caption{\label{fig:2} ({\em Online color}).
{\bf a} Maximal Lyapunov exponent
from direct numerical simulations of the two-dimensional model described in
\Secref{model} (symbols) in the limit $\beta\rightarrow\infty$.
Results from resummation of perturbation theory in $\ku$ \eqnref{lambda_smallKu_2D}
(solid red line) and asymptotic result for large $\ku$ \eqnref{compressible}
(dashed line).  {\bf b} Lyapunov exponents for $\ku=100$ as a function of
$\beta^2$. Numerical results: $\lambda_1$ ($\circ$) and $\lambda_2$ ($\Box$), theoretical results,
Eqs.~(\ref{eq:lambda_largeKu_2D},\ref{eq:sum_lambda_largeKu_2D}), solid lines. }
\end{figure}
Eqs.~(\ref{eq:lambda_largeKu_2D}) and (\ref{eq:sum_lambda_largeKu_2D}) give estimates for the Lyapunov exponents
for large but finite values of $\ku$, as opposed to Eqs.~(\ref{eq:lambda_smallKu_2D})-(\ref{eq:sum_lambda_smallKu_2D})
that give the corresponding expressions for small values of $\ku$.

Let us consider the purely compressible limit  $\beta\to\infty$
in (\ref{eq:lambda_largeKu_2D}) and (\ref{eq:sum_lambda_largeKu_2D}):
\begin{align}
\label{eq:compressible}
\lambda_{1,2}&\sim\frac{1}{\sqrt{12\pi}}\ku\Big[-8\sqrt{2}\pm(5\sqrt{2}-3\arctan(\sqrt{2}))\Big]\,,\\
\lambda_{1}+\lambda_{2}&\sim-\frac{16}{\sqrt{6\pi}}\ku\,.
\label{eq:sum_compressible}
\end{align}
Both the maximal exponent $\lambda_1$ and the sum $\lambda_1+\lambda_2$ are negative in this limit, as expected:
the particles converge to the minima of $V(\ve x,t)=-\psi(\ve x,t)$ at large Kubo numbers.
Eqs.~(\ref{eq:compressible}) and (\ref{eq:sum_compressible}) predict that the Lyapunov
exponents scale as $\ku$ for large values of $\ku$. At small values of $\ku$, by contrast,
the scaling is $\ku^2$ as Eqs.~(\ref{eq:lambda_smallKu_2D}) and (\ref{eq:sum_lambda_smallKu_2D}) show.
Fig.~\ref{fig:2}{\bf a} shows the asymptotic result (\ref{eq:compressible}) for $\lambda_1$
in comparison with results of numerical simulations for particles suspended in a two-dimensional
compressible (purely potential) velocity field. We observe good agreement.

We expect Eqs.~(\ref{eq:lambda_largeKu_2D}) and (\ref{eq:sum_lambda_largeKu_2D}) to give reliable estimates
when the particles are typically found very close to the minima of the potential.
Fig.~\ref{fig:2}{\bf b} shows numerical results for the Lyapunov exponents at $\ku=100$ in partially compressible flows
as a function of $\beta^2$, compared with Eqs.~(\ref{eq:lambda_largeKu_2D}) and (\ref{eq:sum_lambda_largeKu_2D}).
We observe that Eqs.~(\ref{eq:lambda_largeKu_2D}) and (\ref{eq:sum_lambda_largeKu_2D}) yield reasonable estimates
even for small degrees of compressibility. We note that the theory must fail in the incompressible limit ($\beta=0$),
and is only an approximation for finite values of $\beta$. The results show, however, that the dynamics is dominated
by the attracting stagnation points of the velocity field at large Kubo numbers.

\section{Conclusions}
\label{sec:conc}
In this paper we have computed the Lyapunov exponents of small tracer particles advected in one- and two-dimensional
compressible random velocity fields at finite Kubo numbers. For small Kubo numbers we have obtained results by Pad\'e{}-Borel resummation
of perturbation expansions in $\ku$. For large Kubo numbers we have computed the Lyapunov exponents using the Kac-Rice formula.
At finite Kubo numbers the resulting Lyapunov exponents are determined by the details the velocity-field
fluctuations (at small values of $\ku$ by the Eulerian $n$-point functions of the velocity field and its derivatives,
and at large values of $\ku$ by the statistics of its stagnation points).
Our results generalise earlier results \cite{LeJ85,Che98,Fal01,Bec04,Wil07} for compressible flows obtained in the limit $\ku\rightarrow 0$ to
finite Kubo numbers. We find that $\lambda \sim \ku^2$ as $\ku\rightarrow 0$ and $\lambda \sim \ku$ at large (but finite) values of $\ku$.
We have demonstrated that the analytical results are in good agreement with results of numerical simulations, and provide
accurate estimates of the location of the path-coalescence transition for particles advected in compressible flows with finite
Kubo numbers.


The limit $\beta \rightarrow 0$ at large Kubo numbers remains to be analysed. For small values of $\beta$ 
the stagnation points determining the Lyapunov exponents attract only weakly because the corresponding potential minima are shallow:
$\ku$ must be very large for the minima not to disappear before particles are attracted.
In this limit a fraction of particles spends an appreciable amount of time on close-to closed orbits. 
This contribution is not accounted for in the derivation of Eqs.~\eqnref{lambda_largeKu_2D} and \eqnref{sum_lambda_largeKu_2D}. 
For this reason these results are approximate, unless $\beta$ is infinity. 
At $\beta=0$ and $\ku=\infty$ the two-dimensional dynamics (\ref{eq:eqm_r}) corresponds to a one-dimensional
Hamiltonian system. In this case the Lyapunov exponents must vanish. It would be interesting (but outside the scope of this paper)
to consider, if possible, an expansion around this steady case.

We conclude by commenting on two further implications of our results.
First, as mentioned in the introduction, spatial clustering of weakly inertial particles in incompressible velocity fields is often
described in terms of a model where the particles are advected in a \lq synthetic\rq{} velocity field
with a small compressible component (due to the particle inertia). It is known that this approach must
fail when the inertia is large (because of the formation of caustics). But there is also a problem
in the small-inertia limit. Consider the sum of the Lyapunov exponents for inertial particles in
a Gaussian random flow at finite Kubo numbers, Eq.~(9) in Ref.~\cite{Gus11}:
\begin{equation}
\label{eq:st}
\lambda_1 + \lambda_2 = -6 \ku^4 \frac{\st^2(1+3 \st + \st^2)}{(1+\st)^3} + \ldots
\end{equation}
Here $\st$ is the \lq Stokes number\rq{} characterising the importance of particle inertia.
The limit $\st=0$ corresponds to advective dynamics. Comparing Eq.~(\ref{eq:st})
with the leading-order term of Eq.~(\ref{eq:sum_lambda_smallKu_2D}) at small $\st$ and
$\beta$ would lead us to conclude that weakly inertial particles are described
by an advective model with \lq effective compressibility\rq{} $\beta = \sqrt{3/2}\,     \ku \st$.
But this does not give the correct result for $\lambda_1$ (c.f. Eq.~(8) in Ref.~\cite{Gus11}), neither does this correspondence
yield consistent results for terms of higher order in $\ku$ (determined by higher-order
correlation functions of the velocity field). This shows that care is required when
approximating the dynamics of weakly inertial particles by advection in a weakly compressible
velocity field: in general the statistics obtained by sampling along particle trajectories with actual inertial velocities, and in the 
effective compressible velocity field are different.

Second, a related $\ku$-expansion was recently used to compute the tumbling rate of small
axisymmetric particles in three-dimensional random velocity fields at finite Kubo numbers \cite{Gus13a}.
It turns out that the resummation works well also for the series expansion of the tumbling rate.

\begin{acknowledgements} Financial support by Vetenskapsr\aa{}det and by the G\"oran Gustafsson Foundation for Research in Natural Sciences and Medicine and by the EU COST
Action MP0806 on \lq Particles in Turbulence' is gratefullly acknowledged.
The numerical computations were performed using resources provided by C3SE and SNIC.
\end{acknowledgements}

\bibliographystyle{unsrt}      

\end{document}